\documentclass[reprint,aps,11pt, onecolumn, superscriptaddress, floatfix, nofootinbib, amsmath, amssymb, preprintnumbers]{revtex4-2}

\usepackage{dcolumn}
\usepackage{verbatim}

\usepackage{breakurl}

\usepackage{lmodern}

\pdfoutput=1
\usepackage{graphicx}
\usepackage{bm}
\usepackage{amsmath}
\usepackage{amsfonts}
\usepackage{amssymb}
\usepackage{multirow}
\usepackage[colorlinks,linktocpage,linkcolor=cyan,citecolor=cyan]{hyperref}
\usepackage{adjustbox}
\usepackage{array}
\usepackage[capitalise]{cleveref}
\usepackage{acro}
\usepackage[utf8]{inputenc}
\usepackage{CJKutf8}
\usepackage{amssymb}
\usepackage{subfigure}

\usepackage{tikz}
\usepackage{pgfplots}
\pgfplotsset{compat=newest,every axis plot/.append style={line width=1pt}}

\crefname{figure}{Fig.}{Figs.}
\Crefname{figure}{Fig.}{Figs.}
\def\({\left(}
\def\){\right)}
\def\[{\left[}
\def\]{\right]}

\newcommand{\be}{{\begin{eqnarray}}}
\newcommand{\ee}{{\end{eqnarray}}}

\newcommand{\overbar}[1]{\mkern 1.5mu\overline{\mkern-1.5mu#1\mkern-1.5mu}\mkern 1.5mu}

\newcommand{\cH}{\mathcal{H}}
\newcommand{\cS}{\mathcal{S}}
\newcommand{\cO}{\mathcal{O}}

\newcommand{\bk}{\mathbf{k}}
\newcommand{\bq}{\mathbf{q}}

\newcommand{\bx}{\mathbf{x}}

\newcommand{\ud}{\mathrm{d}}

\newcommand{\Beq}{\begin{align}}
\newcommand{\Eeq}{\end{align}}

\DeclareAcronym{BH}{
  short = BH ,
  long = black hole ,
  short-plural = s ,
}

\DeclareAcronym{SNR}{
  short = SNR ,
  long = signal-to-noise ratio ,
  short-plural = s ,
}
\DeclareAcronym{IMRPPv2}{
  short = ,
  long = {\normalsize IMRP}{\footnotesize HENOM}{\normalsize P}v2 ,
  short-plural = ,
}

\DeclareAcronym{SFR}{
  short = SFR ,
  long = star formation rate ,
  short-plural =  ,
}

\DeclareAcronym{IMR}{
  short = IMR ,
  long = inspiral-merger-ringdown ,
  short-plural =  ,
}

\DeclareAcronym{ABH}{
	short = ABH ,
	long  = astrophysical black hole,
  short-plural = s ,
}

\DeclareAcronym{GW}{
  short = GW ,
  long = gravitational wave ,
  short-plural = s ,
}
\DeclareAcronym{SGWB}{
  short = SGWB ,
  long = stochastic gravitational-wave background ,
  short-plural = s ,
}
\DeclareAcronym{CBC}{
  short = CBC ,
  long = compact binary coalescence ,
  short-plural = s ,
}
\DeclareAcronym{BBH}{
  short = BBH ,
  long = binary black hole ,
  short-plural = s ,
}
\DeclareAcronym{PBH}{
  short = PBH ,
  long = primordial black hole ,
  short-plural = s ,
}
\DeclareAcronym{LIGO}{
  short =LIGO ,
  long = Laser Interferometer Gravitational-Wave Observatory ,
  short-plural = ,
}
\DeclareAcronym{LVK}{
  short = LVK ,
  long = {Advanced LIGO, Virgo and KAGRA} ,
  short-plural = ,
}
\DeclareAcronym{ET}{
	short = ET ,
	long  = Einstein Telescope,
  short-plural =  ,
}
\DeclareAcronym{CE}{
	short = CE ,
	long  = Cosmic Explorer,
  short-plural =  ,
}
\DeclareAcronym{LISA}{
	short = LISA ,
	long  = Laser Interferometer Space Antenna,
  short-plural =  ,
}
\DeclareAcronym{BBO}{
	short = BBO ,
	long  = big bang observer,
  short-plural =  ,
}
\DeclareAcronym{DECIGO}{
	short = DECIGO ,
	long  = Deci-hertz Interferometer Gravitational wave Observatory,
  short-plural =  ,
}
\DeclareAcronym{PTA}{
  short = PTA ,
  long = pulsar timing array ,
  short-plural = s ,
}
\DeclareAcronym{FRW}{
  short = FRW ,
  long = Friedman-Robertson-Walker ,
  short-plural =  ,
}
\DeclareAcronym{CMB}{
  short = CMB ,
  long = cosmic microwave background ,
  short-plural =  ,
}

\begin{document}

\title{Primordial Gravitational Waves Assisted by Cosmological Scalar Perturbations}

\author{Yan-Heng Yu}
\affiliation{Theoretical Physics Division, Institute of High Energy Physics, Chinese Academy of Sciences, Beijing 100049, the People's Republic of China}
\affiliation{School of Physical Sciences, University of Chinese Academy of Sciences, Beijing 100049, the People's Republic of China}

\author{Sai Wang}
\email{Corresponding author: wangsai@ihep.ac.cn}
\affiliation{Theoretical Physics Division, Institute of High Energy Physics, Chinese Academy of Sciences, Beijing 100049, the People's Republic of China}
\affiliation{School of Physical Sciences, University of Chinese Academy of Sciences, Beijing 100049, the People's Republic of China}

\begin{abstract} 
Primordial gravitational waves are a crucial prediction of inflation theory, and their detection through their imprints on the cosmic microwave background is actively being pursued. However, these attempts have not yet been successful. In this paper, we propose a novel approach to detect primordial gravitational waves by searching for a signal of second-order tensor perturbations. These perturbations were produced due to nonlinear couplings between the linear tensor and scalar perturbations in the early universe. We anticipate a blue-tilted tensor spectral index, and suggest that the tensor-to-scalar ratio can potentially be measured with high precision using a detector network composed of the ground-based Einstein Telescope and the space-borne LISA project on a decade timescale. 
\end{abstract}

\maketitle

\acresetall

{\emph{Motivation.}}
Primordial gravitational waves, originated from quantized tensor modes of perturbed metric in the very early universe, are one of the most important predictions of cosmic inflation theory \cite{Starobinsky:1979ty,Starobinsky:1980te,Guth:1980zm,Sato:1980yn,1982PhLB..108..389L,Albrecht:1982wi}. On large scales comparable to the whole scale of observable universe, imprints of primordial tensor perturbations on the \ac{CMB} have been proposed before two decades \cite{Seljak:1996gy,Zaldarriaga:1996xe,Kamionkowski:1996zd,Kamionkowski:1996ks}, but have not been observed yet. Recent studies have established upper limits on the spectral amplitude of primordial tensor perturbations \cite{BICEP:2021xfz,Tristram:2020wbi,Beck:2022efr,Campeti:2022vom,Tristram:2021tvh}. The tensor-to-scalar ratio has been shown to be less than 0.032 at the 95\% confidence level, based on precise measurements of anisotropies and polarization in the \ac{CMB} by the Planck satellite and BICEP/Keck Array \cite{Tristram:2021tvh}.

Efforts have been made to detect primordial tensor perturbations on small scales, which are detectable by space-borne and ground-based gravitational-wave interferometers \cite{Liu:2015psa,Huang:2015gka,Meerburg:2015zua,Lasky:2015lej,Cabass:2015jwe,Wang:2016tbj,Berbig:2023yyy}. However, models of canonical single-field slow-roll inflation predict a red-tilted tensor spectrum, with the spectral index exhibiting a consistency relation of $n_t=-r/8$ \cite{Liddle:1992wi}. This makes it particularly challenging for these detectors to measure such a spectrum. Further, a blue-tilted tensor spectrum would imply a violation of the null-energy condition in the effective field theory of single-field inflation models \cite{Creminelli:2006xe,Creminelli:2014wna,Rubakov:2014jja}. To generate a blue-tilted tensor spectrum, additional assumptions, such as higher-derivative operators \cite{Baumann:2015xxa} and strong deviations from single-field slow-roll \cite{Vagnozzi:2020gtf,Benetti:2021uea}, are necessarily involved.

Considering the absence of measurements of primordial tensor perturbations on large scales and the difficulties in generating a blue-tilted tensor spectrum on small scales, it is important to give serious consideration to any new mechanisms that can enhance the tensor spectral amplitude without requiring extraordinary assumptions.

Our proposal suggests that during the early universe, the linear scalar perturbations could have modulated the primordial tensor perturbations, resulting in the production of second-order tensor perturbations with a significantly blue-tilted power spectrum. This anticipated signal can potentially be detected by ongoing and planned ground-based detectors such as the \ac{LVK} \cite{Harry_2010,VIRGO:2014yos,Somiya:2011np}, \ac{ET} \cite{Hild:2010id} and \ac{CE} \cite{Reitze:2019iox}.
Furthermore, scalar perturbations are believed to contribute to the formation of \acp{PBH}, which are considered as a viable candidate for dark matter \cite{Sasaki:2018dmp,Carr:2020xqk}. Additionally, they are expected to produce scalar-induced gravitational waves, which can be detected by planned space-borne detectors such as the \ac{LISA} \cite{2019BAAS...51g..77T,Smith:2019wny}, \ac{BBO} \cite{Crowder:2005nr,Smith:2016jqs}, or \ac{DECIGO} \cite{Seto:2001qf,Kawamura:2020pcg}.
If both the modulated primordial and scalar-induced gravitational waves are detected simultaneously, it would provide valuable insights into the mechanism of cosmic inflation and the nature of dark matter.

This paper investigates the theory of second-order tensor perturbations and the possible multi-band measurements of modulated primordial and scalar-induced gravitational waves using a future detector network consisting of the ground-based \ac{ET} and the space-borne \ac{LISA}. The main objective of this study is to achieve a high-precision measurement of the tensor-to-scalar ratio $r$ with an accuracy of $\Delta{r}\sim\mathcal{O}(10^{-4})$, based on a fiducial model with $r=10^{-2}$ and a bumpy scalar power spectrum with amplitude $\mathcal{A}_{\zeta}=10^{-3}$.

{\emph{Primordial tensor perturbations modulated by cosmological scalar perturbations.}} 
The perturbed \acl{FRW} metric in the Newtonian gauge is $\ud s^{2} = a^{2} \{ -( 1+2\phi )\ud\eta^{2} + [ (1-2\phi)\delta_{ij} + h_{ij} +\tilde{h}_{ij}/2 ] \ud x^{i}\ud x^{j} \}$, where $\tilde{h}_{ij}$ denotes the second-order tensor perturbation sourced by the linear scalar perturbation $\phi$, and the linear tensor perturbation $h_{ij}$. The scalar perturbation in Fourier space is given by $\phi_{\bk}(\eta)=(2/3)\zeta_\bk T_{s}(k\eta)$, where $\zeta_{\bk}$ is the initial comoving curvature perturbation with power spectrum $\langle\zeta_{\bk}\zeta_{\bk^\prime}\rangle=(2\pi^{2}/k^{3})\mathcal{P}_s(k)\delta(\bk+\bk^\prime)$, and the scalar transfer function during the radiation-dominated era is $T_{s}(k\eta)=3(\sin{x}/x-\cos{x})/x^{2}$ with $x=k\eta/\sqrt{3}$ \cite{Maggiore:2018sht}. The tensor perturbation in Fourier space is decomposed into two components, i.e., $h_{\bk,ij}=h^{+}_{\bk}\epsilon^+_{\bk,ij}+h^{\times}_{\bk}\epsilon^{\times}_{\bk,ij}$, where the polarization tensors are defined as 
$
\epsilon^+_{\bk,ij}
    = 
    (
        e_i e_j
        - \overbar{e}_i \overbar{e}_j
    )
    /\sqrt{2} 
$
and
$
    \epsilon^\times_{\bk,ij}
    =   (
            e_i \overbar{e}_j
            + \overbar{e}_i e_j
        )
     /\sqrt{2}
$ 
with $e_{i}$ and $\overbar{e}_{i}$ being orthonormal vectors that are transverse to $\bk$. It is given by $h^{\lambda}_{\bk}(\eta)=H^{\lambda}_{\bk}T_{t}(k\eta)\ (\lambda=+,\times)$, where $H^{\lambda}_{\bk}$ is the initial tensor perturbation with the power spectrum $\langle H^{\lambda}_{\bk}H^{\lambda^\prime}_{\bk^\prime} \rangle=(2\pi^{2}/k^{3})\mathcal{P}_{t}(k)\delta^{\lambda\lambda^\prime}\delta(\bk+\bk^\prime)$
and the tensor transfer function is $T_{t}(k\eta)=\sin(k\eta)/(k\eta)$ \cite{Maggiore:2018sht}. Similarly, we decompose the second-order tensor perturbation in Fourier space into two polarization components, and further decompose each component into three terms, i.e., $\tilde{h}_{\bk}^{\lambda}=\tilde{h}_{\bk}^{\lambda}{}^{ss}+\tilde{h}_{\bk}^{\lambda}{}^{st}+\tilde{h}_{\bk}^{\lambda}{}^{tt}$, where the superscripts ${}^{s}$ and ${}^{t}$ stand for contributions from the linear scalar and tensor perturbations, respectively.

Expanding the Einstein field equations up to second order using the \texttt{xPand} \cite{Pitrou:2013hga} package, we derive the equation of motion for the second-order tensor perturbation. The evolution of $\tilde{h}^{\lambda \alpha\beta}_{\bk}$ with $\alpha\beta=ss,st,tt$ is governed by  
\begin{equation}\label{eq:eom}
\ddot{\tilde{h}}_{\bk}^{\lambda}{}^{\alpha\beta}
    + 2 \cH \dot{\tilde{h}}_{\bk}^{\lambda}{}^{\alpha\beta}
    + k^2 \tilde{h}_{\bk}^{\lambda}{}^{\alpha\beta}
    = 4 \cS_{\bk}^{\lambda}{}^{\alpha\beta}\ ,
\end{equation} 
where an overdot denotes a derivative with respect to $\eta$, $\cH=\dot{a}/a$ is the comoving Hubble parameter, and $\cS_{\bk}^{\lambda}{}^{\alpha\beta}$, as formulated in Eqs.~(\ref{eq:Sss}--\ref{eq:Stt}), is the source term for $\tilde{h}^{\lambda \alpha\beta}_{\bk}$.

We solve Eq.~(\ref{eq:eom}) with the Green's function method and obtain $\tilde{h}_{\bk}\propto\int^{\eta}d\tilde{\eta}\sin(k\eta-k\tilde{\eta})[a(\tilde{\eta})/a(\eta)]\mathcal{S}_{\bk}(\tilde{\eta})$ \cite{Espinosa:2018eve,Kohri:2018awv}, where $a(\eta)\propto\eta$ in the radiation-dominated universe. The power spectrum of gravitational waves is defined as the two-point correlation function, i.e.,   
\begin{eqnarray}
    \langle
        \tilde{h}_\bk^{\lambda\alpha\beta}
        \tilde{h}_{\bk'}^{\lambda'\alpha\beta}
    \rangle
    = 
        \frac{2\pi^2}{k^3}
        \mathcal{P}^{\alpha\beta}_{\tilde{h}}(k)\ 
        \delta^{\lambda \lambda'}
    \delta(\bk + \bk')
    \ ,
\end{eqnarray}
where $\langle...\rangle$ denotes the ensemble average. The dimensionless energy-density spectrum of the second-order tensor perturbations, i.e., the energy density per logarithmic frequency normalized with the critical energy density of the early universe, is given by \cite{Inomata:2016rbd} \begin{eqnarray}\label{eqn:energy density spectra definition} \Omega_\mathrm{gw}^{\alpha\beta}(\eta,k) =\frac{1}{24} \left( \frac{k}{\cH} \right) ^2 \overbar{\mathcal{P}^{\alpha\beta}_{\tilde{h}}(\eta, k)} \ , \end{eqnarray} 
where the overbar denotes the oscillation average and the two polarization modes have been summed over. After tedious but straightforward calculations, we obtain \begin{eqnarray}\label{eq:ogw}
\Omega_\mathrm{gw}^{\alpha\beta}(\eta,k)
    =
    \int _0^{\infty} \ud u
    \int _{\lvert 1-u \rvert} ^{\lvert 1+u \rvert} \ud v
    \ 
    \{...\}_{\alpha\beta} 
    \mathcal{P}_{\alpha}(uk)
    \mathcal{P}_{\beta}(vk)\ ,
\end{eqnarray}
where $\{...\}_{\alpha\beta}$ composed of $u$ and $v$ is formulated in Eqs.~(\ref{eq:ss}--\ref{eq:tt}), and the limit $k\eta\rightarrow\infty$ has been used, implying that the tensor perturbations are deeply within the horizon. The total spectrum is $\Omega_{\mathrm{gw}}=\Omega_{\mathrm{gw}}^{ss}+\Omega_{\mathrm{gw}}^{st}+\Omega_{\mathrm{gw}}^{tt}$. Since the energy density of gravitational waves decays as radiation, the present-day physical energy-density spectrum for the second-order tensor perturbations is approximated by \cite{Wang:2019kaf} \begin{eqnarray} h^2\Omega_\mathrm{gw,0}^{\alpha\beta}(k)=h^2\Omega_\mathrm{r,0}\times\Omega_\mathrm{gw}^{\alpha\beta}(\eta,k)\ , \end{eqnarray} where the corresponding one for photons and neutrinos is $h^2\Omega_\mathrm{r,0}=4.15\times 10^{-5}$, with $h$ being the dimensionless Hubble constant \cite{Planck:2018vyg}.

Before delving into the precision of detection, we present a featured asymptotic behavior of $\Omega_{\mathrm{gw}}^{st}$ in the following. In particular, we remind that the scalar power spectrum on large scales follows a power-law with amplitude $\mathcal{A}_{\zeta,0.05}\simeq2.1\times10^{-9}$ and index $n_{s}\simeq0.96$ at the pivot scale $k_{p}=0.05\ \mathrm{Mpc}^{-1}$ \cite{Planck:2018vyg}. However, the formation of primordial black holes necessitates an enhanced scalar spectral amplitude of $\sim10^{-2}$ on small scales (see Ref.~\cite{Green:2020jor} for a review). We model the scalar power spectrum on small scales as a normal distribution of $\ln k$ with mean $k_\zeta$, standard deviation $\sigma_\zeta$ and spectral amplitude $A_\zeta$ at the scale $k_{\zeta}$, i.e., \cite{Balaji:2022dbi}
\begin{eqnarray}\label{eq:psk}
    \mathcal{P}_{s}(k)=\frac{\mathcal{A}_{\zeta}}{\sqrt{2\pi}\sigma_{\zeta}} \exp\big[{-\frac{\ln^{2} (k/k_\zeta)}{2 \sigma^2_{\zeta}}}\big]\ . 
\end{eqnarray}
On the other hand, we assume that the tensor power spectrum follows a sudden-broken power-law distribution of $k$ throughout the entire scale, i.e., 
\begin{eqnarray}\label{eq:ptk}
    \mathcal{P}_{t}(k) = r \mathcal{A}_{\zeta,0.05} \left(\frac{k}{k_{p}}\right)^{n_{t}} \Theta\left(k_{\mathrm{reh}}-k\right)\ ,
\end{eqnarray}
where $r$ and $n_{t}$ represent the tensor-to-scalar ratio and tensor spectral index, respectively, $k_{\mathrm{reh}}$ is the high-frequency end of the spectrum due to reheating at the end of inflation, and $\Theta(x)$ is the Heaviside function with variable $x$. In models of canonical single-field slow-roll inflation, the consistency relation $n_{t}=-r/8$ holds \cite{Liddle:1992wi}. The current upper bound on the tensor-to-scalar ratio is $r<0.032$ at the 95\% confidence level \cite{Tristram:2021tvh}, indicating a slightly red-tilted tensor spectrum. The reheating frequency $f_{\mathrm{reh}}=k_{\mathrm{reh}}/(2\pi)$ is related to the reheating temperature $T_{\mathrm{reh}}$ and the effective number of relativistic degrees of freedom $g_{\ast,\mathrm{reh}}$ during reheating, with $f_\mathrm{reh}\simeq 0.027 \ \mathrm{Hz}\ (T_\mathrm{reh}/10^6\mathrm{GeV})\ (g_\mathrm{\ast,reh}/106.75)^{1/6}$ \cite{Maggiore:2018sht}. Noticing that the contribution from $g_{\ast,\mathrm{reh}}$ may be negligible due to the small value of the power-law index, thus the reheating frequency is approximately determined by the reheating temperature.

\begin{figure}[t]
\includegraphics[width=\linewidth]{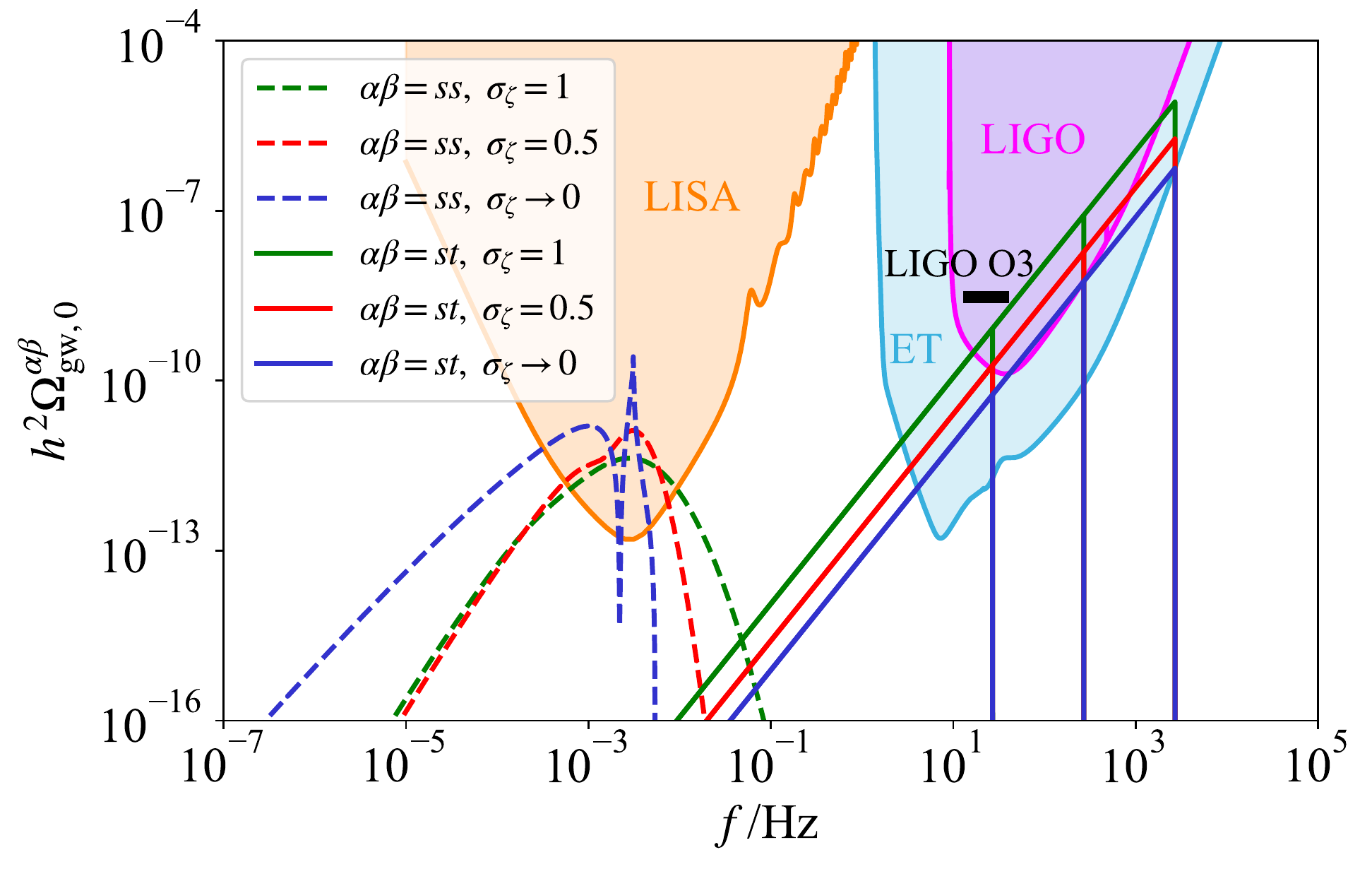}
\caption{Present-day physical energy-density spectra $h^2 \Omega_\mathrm{gw,0}^{ss}$ (dashed lines) and $h^2 \Omega_\mathrm{gw,0}^{st}$ (solid lines) for $\sigma_\zeta \rightarrow 0$ (blue), $\sigma_\zeta =0.5$  (red) and $\sigma_\zeta =1$ (green). The vertical lines from left to right denote $f_\mathrm{reh}=27/270/2700$ Hz. Other parameters are given as $ \mathcal{A}_{\zeta}=10^{-3}$, $f_\zeta=2.7$ mHz, $r=0.01$ and $n_t=-r/8$. The shaded regions show the sensitivities of LISA (orange), LIGO (purple) and ET (blue). The horizontal short line (black) denotes the upper limit of $h^2 \Omega_\mathrm{gw,0}(25 \mathrm{Hz})$ for LIGO O3, using the power-law model marginalizing over the spectral index with a log-uniform prior \cite{KAGRA:2021kbb}.}\label{fig:IGWs and sensitive curves}
\end{figure}

Fig.~\ref{fig:IGWs and sensitive curves} demonstrates that $\Omega_{\mathrm{gw},0}^{st}(k)\propto k^{2+n_{t}}$ as $k_{\zeta}\ll k < k_{\mathrm{reh}}$.
The enhancement results from the leading term $ q^2\phi_{\bk-\bq}h_\bq^{\lambda_1}$ of the source $\cS_\bk^{\lambda st}$ (see Eq.~(\ref{eq:Sst})) in the limit $|\bk-\bq|\ll q\approx k$. On the one hand, for larger momentum $q$ of linear tensor perturbations, the source term $q^2\phi_{\bk-\bq}h_\bq^{\lambda_1}$ can be significantly enhanced by the factor $q^2$. On the other hand, for smaller momentum $|\bk-\bq|$ of linear scalar perturbation, considering $T_s(|\bk-\bq| \eta)\sim 1/(|\bk-\bq| \eta)^2$ within the horizon in the radiation-dominated era, the scalar perturbation decays slower and thus keeps the source term $q^2\phi_{\bk-\bq}h_\bq^{\lambda_1}$ important for a longer time to induce $\Tilde{h}^{\lambda st}_\bk$. 
To make some rough estimates, we have the leading term $\{...\}_{st}\propto 1/u^{4}$ approximately in the limit $u=|\bk-\bq|/k \rightarrow 0$ and $v=|\bq|/k\rightarrow 1$. For simplicity, we take the limit $\sigma_{\zeta}\rightarrow 0$ and get the scalar spectrum $\mathcal{P}_{s}(k)=\mathcal{A}_{\zeta} \delta(\ln(k/k_{\zeta}))$, therefore, the energy-density spectrum can be approximated as $\Omega_{\mathrm{gw}}^{st}(k)\propto \int du \int dv \ u^{-4}\ \delta[\ln(uk/k_{\zeta})]\ k^{n_t}\propto k^{n_t} u^{-2}|_{u=k_\zeta/k} \propto k^{2+n_t}$, where $\int dv$ has been replaced with the integral width $2u$. The spectral index $(2+n_t)$ remains unchanged for different values of $\sigma_{\zeta}$, while the spectral amplitude varies. Further, we can simply use $\Omega_\mathrm{gw}^{st}(k) \simeq \mathrm{few}\times r \mathcal{A}_{\zeta,0.05} \mathcal{A}_{\zeta}  (k/k_{\zeta})^2 (k/k_p)^{n_t} \Theta(k_\mathrm{reh}-k)$ in $k\gg k_\zeta$ region for a good order estimate. The null-energy condition is not violated by this blue-tilted spectrum since second-order gravitational waves were produced during the radiation-dominated era, not the inflationary stage. 

We compare physical energy-density spectra of second-order tensor perturbations (as functions of frequency) with sensitivity curves of LISA, LIGO, and ET in Fig.~\ref{fig:IGWs and sensitive curves}. The scalar-induced tensor perturbations with $\Omega_{\mathrm{gw},0}^{ss}(k)$ have been semi-analytically studied in the literature \cite{Baumann:2007zm,Ananda:2006af,Espinosa:2018eve,Kohri:2018awv}. Due to $r<0.032$, the amplitude of $\Omega_{\mathrm{gw},0}^{tt}(k)$ is too small to fit the scope of Fig.~\ref{fig:IGWs and sensitive curves}. However, the blue-tilted $\Omega_{\mathrm{gw},0}^{st}(k)$ makes it promising to measure primordial tensor perturbations ($r$ and $n_{t}$) and reheating physics ($T_{\mathrm{reh}}$) with high-frequency gravitational-wave detectors. Therefore, we expect that multi-band measurements of second-order tensor perturbations may lead to a better understanding of the late-time stage of inflation.

{\emph{Expected sensitivity of gravitational-wave detectors to measure the anticipated signal.}}
We perform Fisher-matrix forecasts by considering instrumental uncertainties for detector networks composed of space-borne \ac{LISA} and ground-based LIGO or \ac{ET}. The Fisher matrix for second-order tensor perturbations is given by 
\begin{eqnarray}
    F_{ab} =  \sum_{i=1}^{N} T_{i} \epsilon_{i}  \int df \frac{\partial_{\theta_{a}}\Omega_{\mathrm{gw,0}}(k) \ \partial_{\theta_{b}}\Omega_{\mathrm{gw,0}}(k)}{\Omega^{2}_{n,i}(f)} \ ,
\end{eqnarray}
where $f=k/(2\pi)$ is the frequency of gravitational waves, $\theta=\{\ln \mathcal{A}_{\zeta},\sigma_{\zeta},\ln f_{\zeta},r,n_{t},\ln f_{\mathrm{reh}}\}$ is the parameter space being determined, $\Omega_{n}(f)$ denotes the effective detector noise as a function of $f$, as summarized in Ref.~\cite{Campeti:2020xwn}, $N$ is the number of independent detectors, $T$ is the observing time, and $\epsilon$ is the duty circle. For \ac{LISA}, we consider a single detector with 75\% duty circle during a four-year observation. For LIGO (\ac{ET}), we consider two (three) independent detectors with 100\% duty circle during a four-year (one-year) observation. The fiducial parameters are $\mathcal{A}_{\zeta}=10^{-3}$, $\sigma_{\zeta}=0.5$, $f_{\zeta}=2.7$ mHz, $r=0.01$, $n_{t}=-r/8$, and $f_{\mathrm{reh}}=27/270/2700$ Hz. The corresponding spectra have been shown in Fig.~\ref{fig:IGWs and sensitive curves}.

\begin{table}[h]
\begin{tabular}{|c|c|c|c|c|}
    \hline\hline
     Detector  & $f_\mathrm{reh}/\mathrm{Hz}$   & $\Delta r$ & $\Delta n_t$ & $\Delta \ln  f_\mathrm{reh}$ \\
    \hline
     \multirow{3}*{LIGO}  & $27$  & 
     $1.5$ &
     $3.7$ &
     $1.8\times 10^{-3}$
     \\
    \cline{2-5}
     ~  & $270$  & 
     $2.9\times 10^{-2}$ & 
     $6.8\times 10^{-2}$ & 
     $2.2\times 10^{-4}$   \\
    \cline{2-5}
     ~  & $2700$  & 
     $2.0\times 10^{-2}$ & 
     $4.7\times 10^{-2}$ & 
     $2.3\times 10^{-3}$    \\
    \hline
     \multirow{3}*{ET} & $27$  & 
     $1.6\times 10^{-3}$ &
     $3.9\times 10^{-3}$ & 
     $1.9\times 10^{-5}$  \\
    \cline{2-5}
     ~ & $270$  & 
     $3.4\times 10^{-4}$ &
     $6.8\times 10^{-4}$ & 
     $2.6\times 10^{-6}$  \\
    \cline{2-5}
     ~ & $2700$  & 
     $3.0\times 10^{-4}$ &
     $5.4\times 10^{-4}$ & 
     $5.5\times 10^{-5}$  \\
    \hline
\end{tabular}
\caption{The $1\sigma$ confident uncertainties of $r$, $n_t$ and $\ln{f_\mathrm{reh}}$ measured by LIGO and ET for $f_\mathrm{reh}=27/ 270/2700$ Hz.}\label{table:uncertainties}
\end{table}

\begin{figure}[h]
\includegraphics[width=\linewidth]{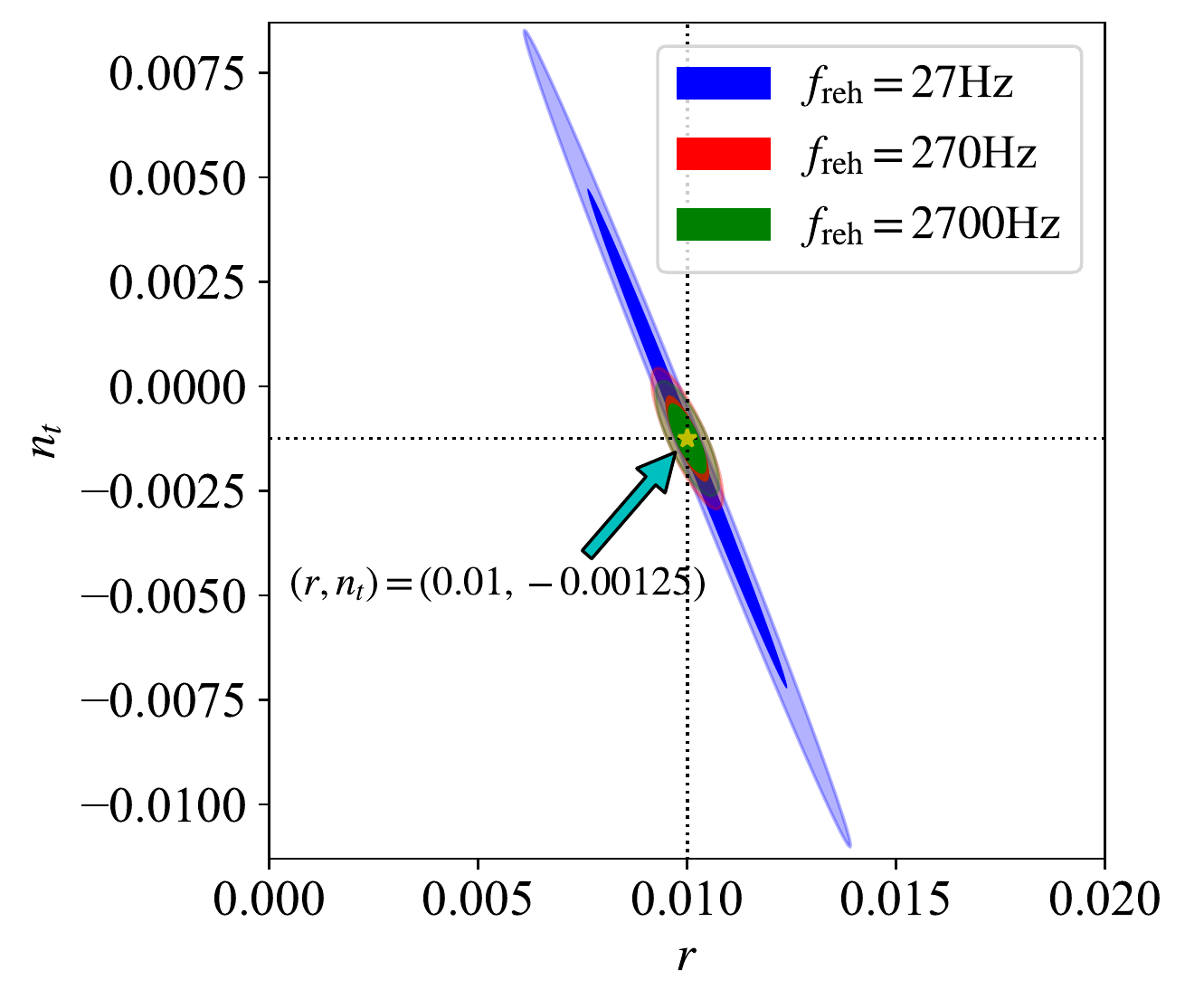}
\caption{Cross-correlations between $r$ and $n_t$ measured by ET for $f_\mathrm{reh}=27(\mathrm{blue})/270(\mathrm{red})/2700(\mathrm{green})$ Hz. Dark and light shaded contours stand for the $1\sigma$ and $2\sigma$ confident regions, respectively. The fiducial model with $r=0.01$ and $n_t=-r/8$ (other parameters are marginalized) is marked as a star.}\label{fig:r-nT uncertainty}
\end{figure}

Though multi-band measurements are performed with detector networks, the parameters of the scalar spectrum in Eq.~(\ref{eq:psk}) are completely determined by \ac{LISA}. The results are given as $\Delta\ln\mathcal{A}_{\zeta}=7.5\times10^{-3}$, $\Delta\sigma_{\zeta}=6.0\times10^{-3}$, and $\Delta\ln f_{\zeta}=3.9\times10^{-3}$, indicating (sub)percent-level measurements. On the other hand, the parameters of the tensor spectrum in Eq.~(\ref{eq:ptk}) are completely determined by LIGO and \ac{ET}. For our fiducial model, LIGO could achieve $\Delta r/r \sim \mathcal{O}(1)$ and $\Delta n_{t}\sim \mathcal{O}(10^{-2})$, while \ac{ET}, with better sensitivity than LIGO, could achieve $\Delta r/r \sim \mathcal{O}(10^{-2})$ and $\Delta n_{t}\sim\mathcal{O}(10^{-4})$, allowing for more-than-$10\sigma$ confident measurements of the tensor-to-scalar ratio and a possibility to test the consistency relation $n_{t}=-r/8$ at the $2\sigma$ confidence level. The precision for measuring $r$ and $n_{t}$ depends on the fiducial value of $f_{\mathrm{reh}}$, as shown in Tab.~\ref{table:uncertainties}. For higher reheating frequency, which implies wider frequency band being captured by LIGO and \ac{ET}, we expect better precision for measurements of $r$ and $n_{t}$. 
Fig.~\ref{fig:r-nT uncertainty} shows the marginalized $1\sigma$ and $2\sigma$ cross-correlations between $r$ and $n_{t}$, as well as their dependence on $f_{\mathrm{reh}}$.  
In addition, the best measurement of $f_{\mathrm{reh}}$ can be performed when $f_{\mathrm{reh}}$ coincides with the most sensitive frequency band of detectors, which is given as $\sim\mathcal{O}(10^{2})$ Hz for LIGO and \ac{ET}. Therefore, we expect the best precision to be $\Delta \ln f_{\mathrm{reh}}\sim\mathcal{O}(10^{-4})$ for LIGO and $\Delta \ln f_{\mathrm{reh}}\sim\mathcal{O}(10^{-6})$ for \ac{ET}. If such a measurement works in the best case, our results may provide meaningful insights for particle physics, as the reheating temperature is $\sim\mathcal{O}(10^{10})$ GeV.

To enhance the detectability of primordial tensor perturbations, our results can be further improved if using fiducial models that anticipate larger amplitudes for $\Omega_{\mathrm{gw},0}^{st}(k)$. This could be achieved, for example, by enhancing the amplitude of the scalar or tensor spectrum, or both, as $\Omega_{\mathrm{gw},0}^{st}\propto r \mathcal{A}_{\zeta}$. In particular, LIGO could potentially measure primordial tensor perturbations by setting the fiducial value to be $\mathcal{A}_{\zeta}\sim10^{-2}$, which is related to an interesting topic of the formation of \acp{PBH} \cite{Green:2020jor}. Other alternatives include increasing the bump width of the scalar spectrum, indicating a larger value for $\sigma_{\zeta}$, or decreasing the peak frequency of the scalar spectrum, indicating a smaller value for $f_{\zeta}$, etc.

{\emph{Conclusion.}}
In the early universe, the linear tensor perturbations were modulated with bump-spectral scalar perturbations to produce second-order tensor perturbations. The resulting tensor spectral index was found to be $(2+n_{t})$, which may have a significant blue tilt. Currently, plans are underway to develop next-generation ground-based gravitational-wave detectors that could provide accurate measurements of the tensor-to-scalar ratio within the next decade. However, such measurements require the existence of both inflationary tensor perturbations and linear scalar perturbations with a bumpy power spectrum, making it difficult to discuss their specifics until the measurements are completed. If future multi-band measurements are able to detect the anticipated signal of second-order tensor perturbations, it could provide valuable insights into the physics of cosmic inflation and help constrain inflation models. While scientists are actively pursuing measurements of \ac{CMB} B-mode polarization (see review in Ref.~\cite{Kamionkowski:2015yta}), our proposal offers an alternative approach to accurately measure primordial tensor perturbations.


\begin{acknowledgements}
We acknowledge Mr. Jun-Peng Li, Dr. Qing-Hua Zhu and Mr. Jing-Zhi Zhou for helpful discussions.
This work is partially supported by the National Natural Science Foundation of China (Grant No. 12175243) and the Key Research Program of the Chinese Academy of Sciences (Grant No. XDPB15). 
 
\end{acknowledgements}


\bibliography{PGW}

\newpage
\appendix

\begin{widetext}

\section{Expression of $\mathcal{S}^{\alpha\beta}_{\bk}$ in Eq.~(\ref{eq:eom})}
\begin{eqnarray}
\cS_{\bk}^\lambda{}^{ss}
     &=&\epsilon^{\lambda,lm}_{\bk}\int \frac{\ud^3 \bq}{(2\pi)^{3/2}}\ 
    q_l q_m
    \Big[
    2\phi_{\bk-\bq}\phi_{\bq}
    +
    \left(\cH^{-1}\dot{\phi}_{\bk-\bq}+\phi_{\bk-\bq}\right)
    \left(\cH^{-1}\dot{\phi}_{\bq}+\phi_{\bq}\right)
    \Big]
    \label{eq:Sss}\ ,\\
\cS_\bk^{\lambda}{}^{st}
    &=& \epsilon^{\lambda,lm}_{\bk} \int \frac{\ud^3 \bq}{(2\pi)^{3/2}}\ 
    \epsilon_{\bq,lm}^{\lambda_1}
    \Big[
    -3\ddot{\phi}_{\bk-\bq} h_{\bq}^{\lambda_1}
    -10\cH \dot{\phi}_{\bk-\bq} h_{\bq}^{\lambda_1}
    -\frac{1}{3}(5k^{2}+5q^{2}-4k^{c}q_{c}) \phi_{\bk-\bq}h_{\bq}^{\lambda_1}
    \Big]
    \label{eq:Sst}\ ,\\
\cS_{\bk}^{\lambda}{}^{tt}&=&
\frac{\epsilon^{\lambda,lm}_{\bk}}{2}
    \int \frac{\ud^3 \bq}{(2\pi)^{3/2}}\ 
    \bigg\{\epsilon^{\lambda_1,b}_{\bk-\bq,l}
    \epsilon^{\lambda_2}_{\bq,bm}
    {k^2}
    \dot{h}_{\bk-\bq}^{\lambda_1}
    \dot{h}_{\bq}^{\lambda_2} 
    + \Big[ \epsilon^{\lambda_1,b}_{\bk-\bq,l}
    \epsilon^{\lambda_2}_{\bq,bm} {(k^{c} q_{c}-q^{2})} 
    - 2 \epsilon^{\lambda_1,bc}_{\bk-\bq}
    \epsilon^{\lambda_2}_{\bq,bm} 
     \ q_c q_l 
  \nonumber\\
    && \quad\quad\quad\quad\quad\quad\quad\quad \      
     -\ \epsilon^{\lambda_1}_{\bk-\bq,mc}
    \epsilon^{\lambda_2}_{\bq,bl} 
     \ (k^b-q^b) q^c  
    - \epsilon^{\lambda_1}_{\bk-\bq,bc}
    \epsilon^{\lambda_2}_{\bq,lm} 
     \ q^b q^c 
     - \frac{1}{2} \epsilon^{\lambda_1,bc}_{\bk-\bq}
    \epsilon^{\lambda_2}_{\bq,bc} 
    \ q^l q^m \Big]
    h_{\bk-\bq}^{\lambda_1}
    h_{\bq}^{\lambda_2}\bigg\}
    \label{eq:Stt}\ .
\end{eqnarray}
\\

\section{Expression of $\{...\}_{\alpha\beta}$ in Eq.~(\ref{eq:ogw})}
\label{app:qi2}
\noindent 
\begin{eqnarray}\label{eq:QI}
&&\big\{...\big\}_{ss} = \frac{3}{1024u^8v^8}\ \Big[{4 v^2-\left(1+v^2-u^2\right)^2}\Big]^{2} 
   \left({u^2+v^2-3}\right)^2   \nonumber\\
   &&\quad\quad\quad\quad
   \times \bigg\{ 
    \left[-4 u v +(u^2+v^2-3)\ln \left| \frac{3-(u+v)^2}{3-(u-v)^2}\right| \ \right]^2
    +\pi^2 (u^2+v^2-3)^2\  \Theta(u+v-\sqrt{3})
    \bigg\}
    \label{eq:ss}\ , \\
&&\big\{...\big\}_{st} = 
     \frac{1}{442368u^8v^8}
    \left[ 16 v^4+24 v^2\left(1+v^2-u^2\right)^2+\left(1+v^2-u^2\right)^4 \right]  \nonumber\\
   &&\quad\quad\quad\quad
   \times \bigg\{ 
    \left[4 u v \left[u^2-9\left( v^2-1\right)\right]+\sqrt{3} \left[u^2-3 \left(v-1\right)^2\right] \left[u^2-3 \left(v+1\right)^2\right] \ln \left| \frac{3-(u+\sqrt{3}v)^2}{3-(u-\sqrt{3}v)^2}\right| \ \right]^2\ \nonumber\\  &&\quad\quad\quad\quad\quad\quad
   \left.+3 \pi^2 \left[u^2-3 \left(v-1\right)^2\right]^2 \left[u^2-3 \left(v+1\right)^2\right]^2\  \Theta\left(u^2 - 3(v - 1)^2\right)
     \right\}
     \label{eq:st}
   \ ,\\
&&\big\{...\big\}_{tt} = 
  \frac{1}{3145728 u^8 v^8 }\left[(u-v)^2-1\right]^2 \left[(u+v)^2-1\right]^2           \nonumber\\  &&\quad\quad\quad\quad\times \bigg\{64 u^2 v^2 \left[u^4+v^4+6 u^2 v^2+6\left(u^2+ v^2\right)+1\right]
  -16 u v \left[5 \left(u^6+v^6\right)+11 u^2 v^2 \left(u^2+v^2\right)\right.
    \nonumber\\    &&\quad\quad\quad\quad\quad\quad
    \left.+11 \left(u^4+v^4\right)-126 u^2 v^2+11 \left(u^2+v^2\right)+5\right]\ln \left|\frac{1-(u+v)^2}{1-(u-v)^2}\right|        \nonumber\\    &&\quad\quad\quad\quad\quad\quad
    +\left[25 \left(u^8+v^8\right)-4 u^2 v^2 \left(u^4+v^4\right)+86 u^4 v^4-4 \left(u^6+v^6\right)+68 u^2 v^2 \left(u^2+v^2\right)\right.
    \nonumber\\    &&\quad\quad\quad\quad\quad\quad \left.+86 \left(u^4+v^4\right)+68 u^2 v^2-4 \left(u^2+v^2\right)+25 \right]
    \left[\pi^2+\ln ^2\left|\frac{1-(u+v)^2}{1-(u-v)^2}\right|\ \right]\bigg\}
    \label{eq:tt}
    \ .
\end{eqnarray}

\section{Supplemental Material}

In this Supplemental Material, we present additional calculations and analysis that complement the main text. The semi-analytical calculation of scalar-induced tensor perturbations are developed in Refs.~\cite{Baumann:2007zm,Ananda:2006af,Espinosa:2018eve,Kohri:2018awv}. The previous studies on ``scalar-tensor" and ``tensor-tensor" mode induced tensor perturbations can be found in Refs.~\cite{Gong:2019mui,Chang:2022vlv}. In our work, for the first time, we provide the semi-analytical expressions for $\Omega_\mathrm{gw}^{st}$ and $\Omega_\mathrm{gw}^{tt}$ as given in Eq.~(\ref{eq:Ogw}). Utilizing these calculations, we propose a novel approach for detecting high-frequency primordial gravitational waves, as discussed in the main text.

In Sec.~\ref{sec:i}, we list the basic equations of second-order tensor perturbations. In Sec.~\ref{sec:ii}, we provide the details of the calculation of $\Omega_\mathrm{gw}^{\alpha\beta}$. In Sec.~\ref{sec:iii}, we analyze the disparities of $\Omega_\mathrm{gw}^{ss}$ and $\Omega_\mathrm{gw}^{st}$ under large-momentum and small-momentum coupling limits, which may be helpful to understand the enhancement of primordial
gravitational waves in the main text.

\subsection{Basic Equations of Second-order Tensor Perturbations}\label{sec:i}

We start with a perturbed spatially-flat Friedman-Robertson-Walker metric in the conformal Newtonian gauge
\begin{eqnarray}\label{eq:frw}
    \ud s^{2}
    = a^{2}(\eta)\ 
    \left\{ 
    -( 1+2\phi )\ \ud\eta^{2}
    + \left[ (1-2\phi)\ \delta_{ij}
    + h_{ij} +\frac{1}{2}\tilde{h}_{ij} \right]
    \ud x^{i}\ud x^{j} \right\}\ ,
\end{eqnarray}
where $a(\eta)$ is the scale factor at the conformal time $\eta$, $\phi$ and $h_{ij}$ denote the linear scalar and tensor perturbations, and $\tilde{h}_{ij}$ denotes second-order tensor perturbations induced by $\phi$ and $h_{ij}$. We expand $\phi$ and $h_{ij}$ ($\tilde{h}_{ij}$ in the same way) in Fourier space
\begin{subequations}\label{eq:phi-Fourier}
\begin{align}
    \phi(\eta,\bx)
    &=
    \int \frac{\ud^3 \bk}{(2\pi)^{3/2}} \ \phi_\bk(\eta) e^{i\bk\cdot\bx} \ ,
    \\
    h_{ij}(\eta,\bx)
    &= \sum_{\lambda=+,\times}
    \int \frac{\ud^3 \bk}{(2\pi)^{3/2}}\  h_\bk^{\lambda}(\eta) \epsilon_{\bk,ij}^{\lambda} e^{i\bk\cdot\bx} \ ,
\end{align}
\end{subequations}
where polarization tensors are 
$\epsilon^+_{\bk,ij}= \left(e_i e_j- \overbar{e}_i\overbar{e}_j\right)/\sqrt{2}$ 
and 
$\epsilon^\times_{\bk,ij}= \left(e_i \overbar{e}_j+ \overbar{e}_i e_j\right)/\sqrt{2}$, 
with orthonormal vectors $e_{i}$ and $\overbar{e}_{i}$ being transverse to the wavevector $\bk$. For adiabatic perturbations, the evolution of the Fourier components  $\phi_\bk$ and $h^{\lambda}_\bk$ are governed by
\begin{subequations}\label{eq:1st eom}
\begin{align}
    \ddot{\phi}_\bk(\eta)
    +3(1+w)\cH\dot{\phi}_\bk (\eta)
    &+w k^2\phi_\bk (\eta)
    =0\ ,
    \\
    \ddot{h}_\bk^{\lambda}(\eta)
    +2\cH\dot{h}_\bk^{\lambda}(\eta)
    &+ k^2 h_\bk^{\lambda}(\eta)
    =0\ ,
\end{align}
\end{subequations}
where an overdot denotes a derivative with respect to $\eta$, $\cH\equiv\dot{a}/a$ is the comoving Hubble parameter, and $w\equiv p/\rho$ is the state parameter with $p$ and $\rho$ being pressure and energy density of the Universe, respectively. Further, the solutions of ~\cref{eq:1st eom} can be written as the primordial curvature (tensor) perturbations $\zeta_\bk$ ($H_\bk^{\lambda}$), times the scalar (tensor) transfer function $T_s$ ($T_t$), i.e.,
\begin{align}\label{eq:transfer functions}
\begin{split}
    \phi_\bk(\eta)
    =\frac{3+3w}{5+3w}T_s(k\eta)\zeta_\bk
    \ ,\ \ 
    h_\bk^{\lambda}(\eta)
    =T_{t}(k\eta)H_\bk^{\lambda}\ .
\end{split}
\end{align}
The dimensionless primordial power spectrum $\mathcal{P}_s(k)$ and $\mathcal{P}_{t}(k)$ are defined as the two-point correlation function, i.e.,   
\begin{eqnarray}\label{eq:primordial power spectrum}
    \langle\zeta_{\bk}\zeta_{\bk^\prime}\rangle=\delta(\bk+\bk^\prime)
    \frac{2\pi^{2}}{k^3}
    \mathcal{P}_s(k)\ ,\ \ 
    \langle H^{\lambda}_{\bk}H^{\lambda^\prime}_{\bk^\prime} \rangle
    =\delta(\bk+\bk^\prime)
    \frac{2\pi^{2}}{k^3}
    \mathcal{P}_{t}(k)\ .
\end{eqnarray}
where $\langle...\rangle$ denotes the ensemble average, the Kronecker symbol $\delta^{\lambda\lambda'}$
and the Dirac function $\delta(\bk+\bk^\prime)$ reflect the independence between two polarizations of the tensor perturbations and the conservation of momentum, respectively. 

Based on the cosmological perturbation theory, each polarization component of the second-order tensor perturbation is composed of three terms, i.e., $\tilde{h}_{\bk}^{\lambda}=\tilde{h}_{\bk}^{\lambda}{}^{ss}+\tilde{h}_{\bk}^{\lambda}{}^{st}+\tilde{h}_{\bk}^{\lambda}{}^{tt}$, where the superscripts ${}^{s}$ and ${}^{t}$ stand for contributions from the linear scalar and tensor perturbations, respectively, and the equation of motion of the second-order gravitational waves  $\tilde{h}^{\lambda \alpha\beta}_{\bk}$ ($\alpha\beta=ss,st,tt$) is given by
\begin{equation}\label{eq:eom}
\ddot{\tilde{h}}_{\bk}^{\lambda}{}^{\alpha\beta}
    + 2 \cH \dot{\tilde{h}}_{\bk}^{\lambda}{}^{\alpha\beta}
    + k^2 \tilde{h}_{\bk}^{\lambda}{}^{\alpha\beta}
    = 4 \cS_{\bk}^{\lambda}{}^{\alpha\beta}\ .
\end{equation} 
The explicit expression of the source term $\cS_{\bk}^{\lambda}{}^{\alpha\beta}$ in Eq.~(\ref{eq:eom}) is obtained to be
\begin{subequations}\label{eq:source1}
\begin{align}
   \cS_{\bk}^\lambda{}^{ss}
     &=\int \frac{\ud^3 \bq}{(2\pi)^{3/2}}\ 
    \epsilon^{\lambda,lm}_{\bk}q_l q_m
    \Big[
    2\phi_{\bk-\bq}\phi_{\bq}
    +\frac{4}{3(1+w)}
    \left(\cH^{-1}\dot{\phi}_{\bk-\bq}+\phi_{\bk-\bq}\right)
    \left(\cH^{-1}\dot{\phi}_{\bq}+\phi_{\bq}\right)
    \Big]\ ,
    \\
    \cS_\bk^{\lambda}{}^{st}
    &=  \int \frac{\ud^3 \bq}{(2\pi)^{3/2}}\ 
    \epsilon^{\lambda,lm}_{\bk}
    \epsilon_{\bq,lm}^{\lambda_1}
    \Big\{
    -3\ddot{\phi}_{\bk-\bq} h_{\bq}^{\lambda_1}
    -2(4+3w)\cH \dot{\phi}_{\bk-\bq} h_{\bq}^{\lambda_1}
    \nonumber\\
    & \quad\quad \ 
    -[(1+2w)(k^2+q^2)-4wk^c q_c]
    \phi_{\bk-\bq}h_{\bq}^{\lambda_1}
    \Big\}\ ,
    \\
\cS_{\bk}^{\lambda}{}^{tt}&=
    \int \frac{\ud^3 \bq}{(2\pi)^{3/2}}\     \frac{\epsilon^{\lambda,lm}_{\bk}}{2}\bigg\{\epsilon^{\lambda_1,b}_{\bk-\bq,l}
    \epsilon^{\lambda_2}_{\bq,bm}
    {k^2}
    \dot{h}_{\bk-\bq}^{\lambda_1}
    \dot{h}_{\bq}^{\lambda_2} 
    + \Big[ \epsilon^{\lambda_1,b}_{\bk-\bq,l}
    \epsilon^{\lambda_2}_{\bq,bm} {(k^{c} q_{c}-q^{2})} 
    - 2 \epsilon^{\lambda_1,bc}_{\bk-\bq}
    \epsilon^{\lambda_2}_{\bq,bm} 
     \ q_c q_l 
  \nonumber\\
    & \quad\quad \      
     -\ \epsilon^{\lambda_1}_{\bk-\bq,mc}
    \epsilon^{\lambda_2}_{\bq,bl} 
     \ (k^b-q^b) q^c  
    - \epsilon^{\lambda_1}_{\bk-\bq,bc}
    \epsilon^{\lambda_2}_{\bq,lm} 
     \ q^b q^c 
     - \frac{1}{2} \epsilon^{\lambda_1,bc}_{\bk-\bq}
    \epsilon^{\lambda_2}_{\bq,bc} 
    \ q^l q^m \Big]
    h_{\bk-\bq}^{\lambda_1}
    h_{\bq}^{\lambda_2}\bigg\}\ .
\end{align}
\end{subequations}
To establish a contact with primordial perturbations during the inflationary stage, we rewrite the source term $\cS_{\bk}^{\lambda}{}^{\alpha\beta}$ above in the form of
\begin{subequations}\label{eq:source2}
\begin{align}
    \cS_{\bk}^{\lambda ss}
    &=
    \int \frac{\ud^3 \bq}{(2\pi)^{3/2}}\ 
    Q_{ss}^{\lambda}(\bk,\bq)
    \ k^2
    f_{ss}(|\bk-\bq|,q,\eta)\ 
    \zeta_{\bk-\bq}\zeta_{\bq}\ ,
    \\
    \cS_{\bk}^{\lambda st}
     &=\int \frac{\ud^3 \bq}{(2\pi)^{3/2}}\ 
     Q_{st}^{\lambda\lambda_1}(\bk,\bq)
     \ k^2
     f_{st}(|\bk-\bq|,q,\eta)\ 
     \zeta_{\bk-\bq} H_\bq^{\lambda_1}\ ,
     \\
     \cS_{\bk}^{\lambda tt}
    &=
    \int \frac{\ud^3 \bq}{(2\pi)^{3/2}}
    \left(\sum_{i=1}^{5}
     Q_{tt,i}^{\lambda \lambda_1 \lambda_2}(\bk,\bq)
     \ k^2
     f_{tt,i}(|\bk-\bq|,q,\eta)
     \right)
     H_{\bk-\bq}^{\lambda_1}H_\bq^{\lambda_2}\ .
\end{align}
\end{subequations}
The projection factor $Q_{\alpha\beta}(\bk,\bq)$ in Eq.~(\ref{eq:source2}) describes the geometric relations between the momenta and polarization tensors of the linear perturbations, being defined as
\begin{subequations}\label{eq:projection factor}
\begin{align}
    Q_{ss}^{\lambda}(\bk,\bq)
    &\equiv
    \epsilon^{\lambda,lm}_\bk\ 
    q_l q_m / k^2\ ,
    \\
    Q_{st}^{\lambda\lambda_1}(\bk,\bq)
   & \equiv
    \epsilon^{\lambda,lm}_\bk\ 
    \epsilon^{\lambda_1}_{\bq,lm}\ ,
    \\
    Q_{tt,1}^{\lambda \lambda_1 \lambda_2}(\bk,\bq)
    &\equiv\ 
    \epsilon^{\lambda,lm}_\bk\ 
    \epsilon^{\lambda_1,b}_{\bk-\bq,l}\ 
    \epsilon^{\lambda_2}_{\bq,bm}\ ,
    \nonumber\\
    Q_{tt,2}^{\lambda \lambda_1 \lambda_2}(\bk,\bq)
    &\equiv-
    \epsilon^{\lambda,lm}_\bk
   \epsilon^{\lambda_1,bc}_{\bk-\bq}\ 
   \epsilon_{\bq,bm}^{\lambda_2}\ 
   (k-q)_l q_c/k^2
   +\Big( (\bk-\bq) \leftrightarrow \bq \ \mathrm{term} \Big)
   \ ,
    \nonumber\\
    Q_{tt,3}^{\lambda \lambda_1 \lambda_2}(\bk,\bq)
    &\equiv\
    \epsilon^{\lambda,lm}_\bk\ 
   \epsilon^{\lambda_1}_{\bk-\bq,mc}\ 
   \epsilon_{\bq,lb}^{\lambda_2}\ 
    k^b k^c/k^2\ ,
    \nonumber\\
    Q_{tt,4}^{\lambda \lambda_1 \lambda_2}(\bk,\bq)
    &\equiv\
   \frac{1}{2}\ 
   \epsilon^{\lambda,lm}_\bk\ 
   \epsilon^{\lambda_1}_{\bk-\bq,bc}\ 
   \epsilon^{\lambda_2}_{\bq,lm}\ 
   k^b k^c/k^2
   +\Big( (\bk-\bq) \leftrightarrow \bq \ \mathrm{term} \Big)
   \ ,
    \nonumber\\
    Q_{tt,5}^{\lambda \lambda_1 \lambda_2}(\bk,\bq)
    &\equiv-
    \epsilon^{\lambda,lm}_\bk\ 
   \epsilon^{\lambda_1,bc}_{\bk-\bq}\ 
   \epsilon_{\bq,bc}^{\lambda_2}\ 
    (k-q)_l q_m/k^2\ .
\end{align}
\end{subequations}
where $Q_{tt,i}^{\lambda \lambda_1 \lambda_2}$ is defined in a symmetric form with respect to $\bk-\bq$ and $\bq$, which facilitates the subsequent manipulation in Eq.~(\ref{eqn:four-point function}) while keeping the integral in Eq.~(\ref{eq:ogw}) unchanged. The source function $f_{\alpha\beta}(|\bk-\bq|,q,\eta)$ in Eq.~(\ref{eq:source2}) describes the time evolution of the linear perturbations, being defined as
\begin{subequations}\label{eq:source function}
\begin{align}
   f_{ss}
    &=\frac{6(1+w)}{5+3w}\ 
    T_s(|\bk-\bq|\eta)\ T_s(q\eta)
   +\frac{12(1+w)}{(5+3w)^2}
   \bigg[\ 
   \frac{1}{\cH}
   \dot{T}_s (|\bk-\bq|\eta)\ T_s (q\eta)
   \nonumber\\
   &\ \ \ 
   +\frac{1}{\cH}
   T_s (|\bk-\bq|\eta)\ 
   \dot{T}_s (q\eta)
   +\frac{1}{\cH^2}
  \dot{T}_s (|\bk-\bq|\eta)\ \dot{T}_s (q\eta)
   \bigg]\ ,
   \\
   f_{st}
   &=\frac{3+3w}{5+3w}
   \ \bigg[
   -\frac{3}{k^2}\ 
   \ddot{T}_s (|\bk-\bq|\eta)\ 
   T_t (q\eta)
   -\frac{2(4+3w)\cH}{k^2}\ 
   \dot{T}_s(|\bk-\bq|\eta)\ 
   T_t(q\eta)
   \nonumber\\
   &\ \ \ 
   -\left(1+\frac{q^2}{k^2}+\frac{2w\ |\bk-\bq|^2}{k^2}\right)\ T_s(|\bk-\bq|\eta)\ 
   T_t(q\eta)
   \bigg]\ ,
   \\
    f_{tt,1}
    &=
    \frac{1}{4}
    \left(1-\frac{|\bk-\bq|^2}{k^2}-\frac{q^2}{k^2}\right)
    T_t(|\bk-\bq|\eta)\ T_t(q\eta)
    +\frac{1}{2k^2}\ 
    \dot{T}_t(|\bk-\bq|\eta)\ 
    \dot{T}_t(q\eta)\ ,
    \nonumber\\
    f_{tt,2}
    &=f_{tt,3}=f_{tt,4}
    =2f_{tt,5}
    =-\frac{1}{2}\ T_t(|\bk-\bq|\eta)\ T_t(q\eta)\ .
\end{align}
\end{subequations}

We can solve Eq.~(\ref{eq:eom}) by Green's function method with the Green's function $G_\bk(\eta,\overbar{\eta})$ being defined as the solution of the equation
\begin{eqnarray}\label{eq:Green equation}
    \frac{\partial^2}{\partial\eta^2}G_\bk(\eta,\overbar{\eta})
    +\left(k^2-\frac{1}{a(\eta)}\frac{\partial^2 a(\eta)}{\partial\eta^2}\right)
    G_\bk(\eta,\overbar{\eta})
    =\delta(\eta,\overbar{\eta})\ ,
\end{eqnarray}
and obtain
\begin{eqnarray}\label{eq:h solution}
    \tilde{h}_{\bk}^{\lambda}{}^{\alpha\beta}(\eta)
    =4 \int^{\eta}d\overbar{\eta} \ 
    \frac{a(\overbar{\eta})}{a(\eta)}
    k G_\bk(\eta,\overbar{\eta})
    \mathcal{S}_{\bk}^{\lambda}{}^{\alpha\beta}(\overbar{\eta})\ .
\end{eqnarray}
Substituting Eq.~(\ref{eq:source2}) into Eq.~(\ref{eq:h solution}), we can recast $\tilde{h}_{\bk}^{\lambda}{}^{\alpha\beta}$ as
\begin{subequations}\label{eq:IGWs}
\begin{align}
   &\tilde{h}_\bk^{\lambda ss}
   =4 \int \frac{\ud^3 \bq}{(2\pi)^{3/2}}\ 
   Q_{ss}^{\lambda}(\bk,\bq)
   I_{ss}(|\bk-\bq|,q,\eta)\ 
   \zeta_{\bk-\bq}\zeta_{\bq}\ ,
   \\
   &\tilde{h}_\bk^{\lambda st}
    =4 \int \frac{\ud^3 \bq}{(2\pi)^{3/2}}\ 
    Q_{st}^{\lambda\lambda_1}(\bk,\bq)
    I_{st}(|\bk-\bq|,q,\eta)\ 
    \zeta_{\bk-\bq} H_\bq^{\lambda_1}\ ,
   \\
   &\tilde{h}_\bk^{\lambda tt}
   =4 \int \frac{\ud^3 \bq}{(2\pi)^{3/2}}
   \left(\sum_{i=1}^{5}
    Q_{tt,i}^{\lambda \lambda_1 \lambda_2}(\bk,\bq)
    I_{tt,i}(|\bk-\bq|,q,\eta)
     \right)
    H_{\bk-\bq}^{\lambda_1}H_\bq^{\lambda_2}\ .
\end{align}
\end{subequations}
where the kernel function $I_{\alpha\beta}(|\bk-\bq|,q,\eta)$ is defined as
\begin{eqnarray}\label{eq:kernel function}
    I_{\alpha\beta}(|\bk-\bq|,q,\eta)
    =\int _0^{\eta}\ud \overbar{\eta}\ 
    \frac{a(\overbar{\eta})}{a(\eta)}
    \ k G_\bk(\eta,\overbar{\eta})
    \ f_{\alpha\beta}(|\bk-\bq|,q,\eta)\ .
\end{eqnarray}
The kernel function $I_{\alpha\beta}(|\bk-\bq|,q,\eta)$ encodes the time evolution of the second-order tensor perturbation $\tilde{h}_\bk^{\lambda \alpha\beta}$, where $a(\overbar{\eta})/a(\eta)$ describes the red-shift effect due to the expansion of the Universe, $kG_\bk(\eta,\overbar{\eta})$ describes the propagation of second-order tensor perturbation, and $f_{\alpha\beta}(|\bk-\bq|,q,\eta)$ describes the evolution of the source terms.

The dimensionless energy-density spectrum of the second-order tensor perturbations, i.e., the energy density per logarithmic frequency normalized with the critical energy density of the early universe, is given by \cite{Inomata:2016rbd} \begin{eqnarray}\label{eq:energy density spectra definition}
      \Omega_\mathrm{gw}^{\alpha\beta}(\eta,k) 
      =\frac{1}{24} 
      \left( \frac{k}{\cH} \right) ^2 \overbar{\mathcal{P}^{\alpha\beta}_{\tilde{h}}(\eta, k)} \ , 
\end{eqnarray} 
where the overbar denotes the oscillation average and the power spectrum of the second-order tensor perturbations $\mathcal{P}^{\alpha\beta}_{\tilde{h}}$ is defined as the two-point correlation function with the two polarization modes being summed over, i.e.,   
\begin{eqnarray}
    \langle
        \tilde{h}_\bk^{\lambda\alpha\beta}
        \tilde{h}_{\bk'}^{\lambda'\alpha\beta}
    \rangle
    = 
    \delta^{\lambda \lambda'}
    \delta(\bk + \bk')\ 
    \frac{2\pi^2}{k^3}\ 
    \mathcal{P}^{\alpha\beta}_{\tilde{h}}(k)
    \ .
\end{eqnarray}
The total spectrum is $\Omega_{\mathrm{gw}}=\Omega_{\mathrm{gw}}^{ss}+\Omega_{\mathrm{gw}}^{st}+\Omega_{\mathrm{gw}}^{tt}$. Since the energy density of tensor perturbations decays as radiation, the present-day physical energy-density spectrum for the second-order tensor perturbations is approximated by \cite{Wang:2019kaf} \begin{eqnarray} h^2\Omega_\mathrm{gw,0}^{\alpha\beta}(k)=h^2\Omega_\mathrm{r,0}\times\Omega_\mathrm{gw}^{\alpha\beta}(\eta,k)\ , \end{eqnarray} where the corresponding one for photons and neutrinos is $h^2\Omega_\mathrm{r,0}=4.15\times 10^{-5}$, with $h$ being the dimensionless Hubble constant \cite{Planck:2018vyg}.

By neglecting the non-Gaussianity of the primordial curvature perturbations, we can use Wick's theorem and get
\begin{subequations}\label{eqn:four-point function}
\begin{align}
   &\Omega_\mathrm{gw}^{ss}
   \propto
   \langle \zeta_{\bk-\bq} \zeta_{\bq} \zeta_{\bk'-\bq'} \zeta_{\bq'} \rangle 
   =
   \Big[
   \delta(\bq+\bq')
   +\delta(\bq+\bk'-\bq')
   \Big]
   \delta(\bk+\bk')\ 
   \frac{2\pi^2\mathcal{P}_s(|\bk-\bq|)}{|\bk-\bq|^3}
   \frac{2\pi^2\mathcal{P}_s(q)}{q^3}\ ,
   \\
   &\Omega_\mathrm{gw}^{st}
   \propto
   \langle \zeta_{\bk-\bq}  H^{\lambda_1}_{\bq}
   \zeta_{\bk'-\bq'}
   H^{\lambda_1'}_{\bq'} \rangle 
   =\delta^{\lambda_1 \lambda_1'}
   \delta(\bq+\bq')
   \delta(\bk+\bk')\ 
   \frac{2\pi^2\mathcal{P}_s(|\bk-\bq|)}{|\bk-\bq|^3}
   \frac{2\pi^2\mathcal{P}_t(q)}{q^3}\ ,
   \\
   &\Omega_\mathrm{gw}^{tt}
   \propto
   \langle H_{\bk-\bq}^{\lambda_1} H_{\bq}^{\lambda_2} H_{\bk'-\bq'}^{\lambda_1'} H_{\bq'}^{\lambda_2'} 
   \rangle
   \nonumber\\
   &\quad\quad
   =
   \left[
   \delta^{\lambda_1 \lambda_1'}\delta^{\lambda_2 \lambda_2'}
   \delta(\bq+\bq')
   +\delta^{\lambda_1 \lambda_2'}\delta^{\lambda_2 \lambda_1'}
   \delta(\bq+\bk'-\bq')
   \right]
   \delta(\bk+\bk')\ 
   \frac{2\pi^2\mathcal{P}_t(|\bk-\bq|)}{|\bk-\bq|^3}
   \frac{2\pi^2\mathcal{P}_t(q)}{q^3}
   \ .
\end{align}
\end{subequations} 
We can finally obtain the general expression of $\Omega_\mathrm{gw}^{\alpha\beta}$ after straightforward calculations, i.e.,
\begin{eqnarray}\label{eq:ogw}
    \Omega_\mathrm{gw}^{\alpha\beta}(\eta,k)
    =
    \int _0^{\infty} \ud u
    \int _{\lvert 1-u \rvert} ^{\lvert 1+u \rvert} \ud v
    \ 
    \frac{k^2}{6\cH^2 u^2v^2}\ 
    Q_{\alpha\beta}^2(u,v)
    \overbar{I_{\alpha\beta}^{2}(u,v,x\rightarrow\infty)}\ 
    \mathcal{P}_{\alpha}(uk)
    \mathcal{P}_{\beta}(vk)\ ,
\end{eqnarray}
where $u\equiv|\bk-\bq|/k$, $v\equiv q/k$, $x\equiv k\eta$ , and the limit $x\rightarrow\infty$ has been used, implying that the tensor perturbations are deeply within the horizon. In Eq.~(\ref{eq:ogw}),  $Q_{ss}^2$ and $Q_{st}^2$ are defined as the sum of the polarizations over two projection factors $Q_{ss}^{\lambda}(\bk,\bq)$ and $Q_{st}^{\lambda\lambda_1}(\bk,\bq)$ in Eq.~(\ref{eq:projection factor}), i.e.,
\begin{subequations}\label{eq:QQssst}
\begin{align}
    Q_{ss}^2(u,v)
    &\equiv    \delta^{\lambda\lambda'}
    Q_{ss}^{\lambda}(\bk,\bq)
    Q_{ss}^{\lambda'}(\bk,\bq)
    =
    \bigg(\frac{4 v^2-\left(1+v^2-u^2\right)^2}{4}\bigg)^2\ ,
    \\
    Q_{st}^2(u,v)
    &\equiv
    \delta^{\lambda\lambda'}
    \delta^{\lambda_1\lambda_1'}
    Q_{st}^{\lambda\lambda_1}(\bk,\bq)
    Q_{st}^{\lambda'\lambda_1'}(\bk,\bq)
    =
    \frac{1}{4}+\frac{3 \left(1+v^2-u^2\right)^2}{8v^2}+\frac{\left(1+v^2-u^2\right)^4}{64v^4} \ ,
\end{align}
\end{subequations}
while a bit differently, $Q_{tt}^2(u,v)$ are defined together with $I_{tt}^2(u,v,x)$ for convenience, given by the expression below composed of $Q_{tt,i}^{\lambda \lambda_1 \lambda_2}$ in Eq.~(\ref{eq:projection factor}) and $I_{tt,i}$ in Eq.~(\ref{eq:IGWs}), i.e.,
\begin{eqnarray}\label{eq:QQIItt definition}
    Q_{tt}^2 \times I_{tt}^2
    \equiv
    \delta^{\lambda \lambda'}
    \left(
    \delta^{\lambda_1 \lambda_1'}
     \delta^{\lambda_2 \lambda_2'}
    + \delta^{\lambda_1 \lambda_2'}
     \delta^{\lambda_2 \lambda_1'}
     \right)
     \bigg(
     \sum_{i=1}^{5}\
    Q_{tt,i}^{\lambda \lambda_1 \lambda_2}
    I_{tt,i}
    \bigg)
    \bigg(
    \sum_{j=1}^{5}\
    Q_{tt,j}^{\lambda' \lambda_1' \lambda_2'}
    I_{tt,j}
    \bigg)\ .
\end{eqnarray}
where the contractions of polarization indices in Eq.~(\ref{eq:QQIItt definition}) correspond to those in Eq.~(\ref{eqn:four-point function}). Remind that $Q_{tt,i}$ in Eq.~(\ref{eq:projection factor}) and $f_{tt,i}$ in Eq.~(\ref{eq:source function}) are already symmetric about $\bk-\bq$ and $\bq$, so the two types of momentum contractions in $\Omega_\mathrm{gw}^{tt}$ in Eq.~(\ref{eqn:four-point function}) yield the same result.

It is also important to mention that we have substituted the variables $|\bk-\bq|$, $q$, and $\eta$ with corresponding dimensionless variables rescaled by $k$ (i.e., $u\equiv |\bk-\bq|/k $, $v\equiv q/k$, and $x\equiv k\eta$) in Eq.~(\ref{eq:ogw}) and subsequent calculations. However, for simplicity, we still use the original names of the functions (e.g., $I_{\alpha\beta}(|\bk-\bq|,q,\eta)$ is substituted by $I_{\alpha\beta}(u,v,x)$).


\subsection{The calculation of energy density spectrum in RD era}\label{sec:ii}

We consider a radiation-dominated (RD) era after inflation and easily get $w=1/3$, $a(\eta)\propto \eta$ and $\cH=1/\eta$. Solving Eq.~(\ref{eq:1st eom}) , the linear perturbations in Eq.~(\ref{eq:transfer functions}) are given by 
$\phi_\bk(\eta)=(2/3)T_s(k\eta)\zeta_\bk$ and $h^\lambda_\bk=T_t(k\eta)H^\lambda_\bk$ with the transfer functions being
\begin{subequations}\label{eq:transfer functions expression}
\begin{align}
    T_{s}(x)
    &=\frac{9}{x^2}
    \left(
    \frac{\sqrt{3}}{x}\sin{\frac{x}{\sqrt{3}}}
    -\cos{\frac{x}{\sqrt{3}}}
    \right)\ ,
    \\
    T_t(x)
    &=\frac{\sin{x}}{x}\ .
\end{align}
\end{subequations}
Substituting Eq.~(\ref{eq:transfer functions expression}) into Eq.~(\ref{eq:source function}), the source functions in RD era are given by
\begin{subequations}\label{eq:source function expressions}
\begin{align}
    f_{ss}(u,v,x)
    &=\frac{12}{u^2 v^2 x^6}\ 
    \bigg\{
    18 uvx^2 
    \cos{\frac{ux}{\sqrt{3}}}
    \cos{\frac{vx}{\sqrt{3}}}
    +[\ 54-6(u^2+v^2)x^2+u^2 v^2 x^4\ ]
    \sin{\frac{ux}{\sqrt{3}}}
    \sin{\frac{vx}{\sqrt{3}}}
    \nonumber\\
    &\quad\quad\quad
    +2\sqrt{3}\ ux(v^2 x^2-9)
    \cos{\frac{ux}{\sqrt{3}}}
    \sin{\frac{vx}{\sqrt{3}}}
    +2\sqrt{3}\ vx(u^2 x^2-9)
    \sin{\frac{ux}{\sqrt{3}}}
    \cos{\frac{vx}{\sqrt{3}}}
    \bigg\}
    \ ,
    \\
    f_{st}(u,v,x)
    &=
    -\frac{1}{3 u^3 v x^6}\ 
     \bigg\{
     \left[\  ux^3 \left(u^2-3 v^2-3\right)-18\ u x\  \right] 
    \cos {\frac{u x}{\sqrt{3}}}\ \sin {v x}
    \nonumber\\
    &\quad\quad\quad
    + 
    \left[\ 3 \sqrt{3}x^2 \left(-u^2+v^2+1\right)+18 \sqrt{3}\ \right] 
    \sin {\frac{u x}{\sqrt{3}}}\ \sin {v x}
    \bigg\}\ ,
    \\
    f_{tt,1}(u,v,x)
    &=
    -\frac{\left(u^2+v^2-3\right)
    \  \sin {u x}
    \  \sin {v x}}
    {4 u v x^2}\ ,
    \nonumber\\
    f_{tt,2}(u,v,x)
    &=f_{tt,3}(u,v,x)=f_{tt,4}(u,v,x)=2f_{tt,5}(u,v,x)
    =-\frac{\sin{ux}\ \sin{vx}}{2uvx^2}\ .
\end{align}
\end{subequations}
On the other hand, the solution of Eq.~(\ref{eq:Green equation}) with $a(\eta)\propto\eta$ is
\begin{eqnarray}\label{eq:Green function}
    G_\bk(x,\overbar{x})
    =\sin{(x-\overbar{x})}/k
\end{eqnarray}

Substuting Eq.~(\ref{eq:source function expressions}) and Eq.~(\ref{eq:Green function}) into Eq.~(\ref{eq:kernel function}), we obtain the explicit expression of the kernel function $I_{\alpha\beta}(u,v,x)$, in which we have used $a(\overbar{\eta})/a(\eta)\simeq \overbar{\eta}/\eta$ in RD era by neglecting the tiny correction from the change of the relativistic degrees of freedom, i.e.,
\begin{subequations}\label{eq:I}
\begin{align}
   &I_{ss}(u,v,x)
   =\ \frac{3\cos{x}}{4u^3v^3 x}
   \ \bigg\{ 
   (u^2+v^2-3)^2
   \bigg[ 
   -\mathrm{Si} \left( \left( 1-\frac{v-u}{\sqrt{3}} \right) x \right)
   \nonumber\\
   &\quad\quad
   +\mathrm{Si} \left( \left( 1-\frac{v+u}{\sqrt{3}} \right) x \right)
   -\mathrm{Si} \left( \left( 1+\frac{v-u}{\sqrt{3}} \right) x \right)
   +\mathrm{Si} \left( \left( 1+\frac{v+u}{\sqrt{3}} \right) x \right)
   \bigg]
   \bigg\}
   \nonumber\\
   &\quad\quad
   -\frac{3\sin{x}}{4u^3v^3 x}
   \bigg\{
   4u v(u^2+v^2-3)
   -(u^2+v^2-3)^2
   \bigg[
   \mathrm{Ci} \left( \left( 1-\frac{v-u}{\sqrt{3}} \right) x \right)
   -\mathrm{Ci} \left( \left| 1- \frac{v+u}{\sqrt{3}} \right| x \right)
   \nonumber\\
   &\quad\quad
   +\mathrm{Ci} \left( \left( 1+\frac{v-u}{\sqrt{3}} \right) x \right)
   -\mathrm{Ci} \left( \left( 1+\frac{v+u}{\sqrt{3}} \right) x \right)
   +\ln \left(\left| \frac{3-(u+v)^2}{3-(u-v)^2}\right| \right)
   \bigg]
   \bigg\}
   +\cO\left( \frac{1}{x^2} \right)
   \ ,
   \nonumber\\
   \\
   &I_{st}(u,v,x)
   =\frac{\cos{x}}{24u^3vx}
   \bigg\{ 
   \sqrt{3} 
   \left[u^2-3 \left(v-1\right)^2\right] \left[u^2-3 \left(v+1\right)^2\right]
   \bigg[\ 
   \mathrm{Si} \left( \left( 1-\frac{u}{\sqrt{3}}+v  \right) x \right)
   \nonumber\\
   &\quad\quad
   -\mathrm{Si} \left( \left( 1- \frac{u}{\sqrt{3}}-v \right) x \right)
   -\mathrm{Si} \left( \left( 1+ \frac{u}{\sqrt{3}}+v \right) x \right)
   +\mathrm{Si} \left( \left( 1+ \frac{u}{\sqrt{3}}-v \right) x \right)
   \bigg]
   \bigg\}
   \nonumber\\
   &\quad\quad
   +\frac{\sin{x}}{24u^3vx}
   \bigg\{ 
   4 u v \left[u^2-9\left( v^2-1\right)\right]+
   \sqrt{3} 
   \left[u^2-3 \left(v-1\right)^2\right] \left[u^2-3 \left(v+1\right)^2\right]
   \nonumber\\
   &\quad\quad
   \times
   \bigg[
   \ 
   \mathrm{Ci} \left( \left( 1- \frac{u}{\sqrt{3}}+v \right) x \right)
   -\mathrm{Ci} \left( \bigg\lvert 1- \frac{u}{\sqrt{3}}-v  \bigg\rvert x \right)
   -\mathrm{Ci} \left( \left( 1+ \frac{u}{\sqrt{3}}+v \right) x \right)
   \nonumber\\
   &\quad\quad
   +\mathrm{Ci} \left( \bigg\lvert 1+ \frac{u}{\sqrt{3}}-v \bigg\rvert x \right)
   +\ln \left| \frac{3-(u+\sqrt{3}\ v)^2}{3-(u-\sqrt{3}\ v)^2}\right|
   \bigg]
   \bigg\}
   +\cO\left( \frac{1}{x^2} \right)
   \ ,
   \nonumber\\
   \\
   &I_{tt,1}(u,v,x)
   =\ -\frac{\sin{x}}{4x}+\cO\left( \frac{1}{x^2} \right)\ ,
   \nonumber\\
   &I_{tt,2}(u,v,x)
   =\ I_{tt,3}(u,v,x)
   =\ I_{tt,4}(u,v,x)
   =2 I_{tt,5}(u,v,x)
   \nonumber\\
   &\quad
   =
   \frac{\cos{x}}{8uvx}
   \bigg\{
   \mathrm{Si}\Big((1-u+v)x\Big)
   -\mathrm{Si}\Big((1-u-v)x\Big)
   +\mathrm{Si}\Big((1+u-v)x\Big)
   -\mathrm{Si}\Big((1+u+v)x\Big)
   \bigg\}
   \nonumber\\
   &\quad\quad
   -\frac{\sin{x}}{8uvx}\ 
   \bigg\{ 
   \mathrm{Ci}\Big((1-u+v)x\Big)
   +\mathrm{Ci}\Big((1+u-v)x\Big)
   -\mathrm{Ci}\Big(\left|1-u-v\right|x\Big)
   -\mathrm{Ci}\Big((1+u+v)x\Big)
   \nonumber\\
   &\quad\quad
   +\ln\left|{\frac{1-(u+v)^2}{1-(u-v)^2}}\right|\ 
   \bigg\}\ 
   +\cO\left( \frac{1}{x^2} \right)\ .
\end{align}
\end{subequations}
where $\mathrm{Si}(x)$ and $\mathrm{Ci}(x)$ are defined as $\mathrm{Si}(x)\equiv\int_0^x \ud y\ (\sin{y}/y)$
and $\mathrm{Ci}(x)\equiv\ -\int_x^\infty \ud y\ (\cos{y}/y)$.
Thus the square of kernel function $I_{\alpha\beta}(u,v,x)$ in the limit of $ k\eta \rightarrow \infty$ and oscillation average are given by
\begin{subequations}\label{eq:II}
\begin{align}
   \overbar{I^2_{ss}(u,v,x \rightarrow \infty)} \ 
   &=\ \frac{1}{2}
   \left(\frac{3(u^2+v^2-3)}{4u^3v^3 x}\right)^2
   \bigg\{ 
   \pi^2 (u^2+v^2-3)^2\  \Theta(u+v-\sqrt{3})
    \nonumber\\
    &\quad
   +\left[-4 u v +(u^2+v^2-3)
   \ln \left| \frac{3-(u+v)^2}{3-(u-v)^2}\right| \ \right]^2
    \bigg\}\ ,
    \\
    \overbar{I_{st}^2(u,v,x\rightarrow \infty)}
    &=\frac{1}{2}
   \left(\frac{1}{24u^3vx}\right)^2
   \bigg\{ 
   3 \pi^2 \left[u^2-3 \left(v-1\right)^2\right]^2 \left[u^2-3 \left(v+1\right)^2\right]^2 
   \nonumber\\
   &\quad
   \times
   \Theta\left(u^2 - 3(v - 1)^2\right)
   +\bigg[
   4 u v \left[u^2-9\left( v^2-1\right)\right]
   \nonumber\\
   &\quad
   +\sqrt{3} \left[u^2-3 \left(v-1\right)^2\right] \left[u^2-3 \left(v+1\right)^2\right] \ln \left| \frac{3-(u+\sqrt{3}v)^2}{3-(u-\sqrt{3}v)^2}\right| 
   \bigg]^2
   \bigg\}\ ,
\end{align}
\end{subequations}
and the expression of $Q^2_{tt}(u,v)\ \overbar{I_{tt}^2(u,v,x\rightarrow \infty)}$ in Eq.~(\ref{eq:QQIItt definition}) is given by
\begin{align}\label{eq:QQIItt }
    & Q^2_{tt}(u,v)\ 
    \overbar{I_{tt}^2(u,v,x\rightarrow \infty)}
    \nonumber\\
    &
    =\frac{1}{524288 u^6 v^6 x^2}
    \left[(u-v)^2-1\right]^2 \left[(u+v)^2-1\right]^2           \nonumber\\
   &\quad
    \times \bigg\{64 u^2 v^2 \left[u^4+v^4+6 u^2 v^2+6\left(u^2+ v^2\right)+1\right]
    -16 u v \left[5 \left(u^6+v^6\right)
    \right.
    \nonumber\\
   &\quad
    +11 u^2 v^2 \left(u^2+v^2\right)
    \left.+11 \left(u^4+v^4\right)-126 u^2 v^2+11 \left(u^2+v^2\right)+5\right]\ln \left|\frac{1-(u+v)^2}{1-(u-v)^2}\right|        
    \nonumber\\
    &\quad
    +\left[25 \left(u^8+v^8\right)-4 u^2 v^2 \left(u^4+v^4\right)+86 u^4 v^4-4 \left(u^6+v^6\right)+68 u^2 v^2 \left(u^2+v^2\right)\right.
    \nonumber\\
    &\quad
    \left.+86 \left(u^4+v^4\right)+68 u^2 v^2-4 \left(u^2+v^2\right)+25 \right]
    \left[\pi^2+\ln ^2\left|\frac{1-(u+v)^2}{1-(u-v)^2}\right|\ \right]\bigg\} \ .
\end{align}
where we have used the limit $\lim_{x\rightarrow \infty} \mathrm{Si}(Ax)=\mathrm{sgn}(A)\pi/2$ and $\lim_{x\rightarrow \infty}\mathrm{Ci}(Ax)=0$, and the Heaviside function $\Theta$ in Eq.~(\ref{eq:II}) comes from the discussion of the sign of the variable of $\mathrm{Si}$ function.
We finally get the expressions of the fractional energy density spectrum of the second-order tensor perturbations $\Omega_\mathrm{gw}^{\alpha\beta}(k)$ in Eq.~(\ref{eq:ogw}), namely
\begin{subequations}\label{eq:Ogw}
\begin{align}
   \Omega_\mathrm{gw}^{ss}(k)
   & =
   \int _0^{\infty} \ud u
   \int _{\lvert 1-u \rvert} ^{\lvert 1+u \rvert} \ud v
   \ \mathcal{P}_{s}(uk)
    \mathcal{P}_{s}(vk)
    \nonumber\\
   &\quad
    \times
   \frac{3}{1024u^8v^8}\ \Big[{4 v^2-\left(1+v^2-u^2\right)^2}\Big]^{2} 
   \left({u^2+v^2-3}\right)^2   \nonumber\\
   &\quad
   \times \bigg\{ 
    \left[-4 u v +(u^2+v^2-3)\ln \left| \frac{3-(u+v)^2}{3-(u-v)^2}\right| \ \right]^2
    +\pi^2 (u^2+v^2-3)^2\  \Theta(u+v-\sqrt{3})
    \bigg\}\ , 
    \nonumber\\
   \\
    \Omega_\mathrm{gw}^{st}(k)
    & =
   \int _0^{\infty} \ud u
   \int _{\lvert 1-u \rvert} ^{\lvert 1+u \rvert} \ud v
   \ \mathcal{P}_{s}(uk)
    \mathcal{P}_{t}(vk)
    \nonumber\\
   &\quad
    \times
     \frac{1}{442368u^8v^8}
    \left[ 16 v^4+24 v^2\left(1+v^2-u^2\right)^2+\left(1+v^2-u^2\right)^4 \right]  \nonumber\\
   &\quad
   \times \bigg\{ 
    \left[4 u v \left[u^2-9\left( v^2-1\right)\right]+\sqrt{3} \left[u^2-3 \left(v-1\right)^2\right] \left[u^2-3 \left(v+1\right)^2\right] \ln \left| \frac{3-(u+\sqrt{3}v)^2}{3-(u-\sqrt{3}v)^2}\right| \ \right]^2 \nonumber\\
   &\quad
   \left.+3 \pi^2 \left[u^2-3 \left(v-1\right)^2\right]^2 \left[u^2-3 \left(v+1\right)^2\right]^2\  \Theta\left(u^2 - 3(v - 1)^2\right)
     \right\}
   \ ,
   \nonumber\\
   \\
   \Omega_\mathrm{gw}^{tt}(k)
   & =
   \int _0^{\infty} \ud u
   \int _{\lvert 1-u \rvert} ^{\lvert 1+u \rvert} \ud v
   \ \mathcal{P}_{t}(uk)
    \mathcal{P}_{t}(vk)
    \nonumber\\
   &\quad
    \times
    \frac{1}{3145728 u^8 v^8}
    \left[(u-v)^2-1\right]^2 \left[(u+v)^2-1\right]^2           \nonumber\\
   &\quad
    \times \bigg\{64 u^2 v^2 \left[u^4+v^4+6 u^2 v^2+6\left(u^2+ v^2\right)+1\right]
    -16 u v \left[5 \left(u^6+v^6\right)
    \right.
    \nonumber\\
   &\quad
    +11 u^2 v^2 \left(u^2+v^2\right)
    \left.+11 \left(u^4+v^4\right)-126 u^2 v^2+11 \left(u^2+v^2\right)+5\right]\ln \left|\frac{1-(u+v)^2}{1-(u-v)^2}\right|        
    \nonumber\\
    &\quad
    +\left[25 \left(u^8+v^8\right)-4 u^2 v^2 \left(u^4+v^4\right)+86 u^4 v^4-4 \left(u^6+v^6\right)+68 u^2 v^2 \left(u^2+v^2\right)\right.
    \nonumber\\
    &\quad
    \left.+86 \left(u^4+v^4\right)+68 u^2 v^2-4 \left(u^2+v^2\right)+25 \right]
    \left[\pi^2+\ln ^2\left|\frac{1-(u+v)^2}{1-(u-v)^2}\right|\ \right]\bigg\} \ .
    \nonumber\\
\end{align}
\end{subequations}


\subsection{The Disparities of "Scalar-Scalar" and "Scalar-Tensor" Modes under Large-Momentum and Small-Momentum Coupling Limits}\label{sec:iii}

We do some limit analysis in $u=|\bk-\bq|/k \rightarrow 0$ and $v=|\bq|/k\rightarrow 1$
to demonstrate the differences between $\Omega_\mathrm{gw}^{ss}$ and $\Omega_\mathrm{gw}^{st}$, which can provide an explanation for why the large-momentum modes can be enhanced in $\Omega_\mathrm{gw}^{st}$, but not in $\Omega_\mathrm{gw}^{ss}$.

(a) One of the differences is ``projection factors" $Q_{\alpha\beta}$, which describes the geometric relations between the momenta and polarization tensors of the linear perturbations. As shown in Eq.~(\ref{eq:QQssst}), we have $Q^2_{ss}(u,v) \rightarrow 0$ in $u \rightarrow 0, v \rightarrow 1$ limit, implying a suppression on the couplings between small-momentum scalar and large momentum tensor. However, as for ``scalar-tensor" mode in Eq.~(\ref{eq:QQssst}), we have $Q^2_{st}(u,v) \rightarrow 2$ keeping a constant in $u \rightarrow 0, v \rightarrow 1$ limit, which means there is always a non-vanishing ``effective quadrupole moment" in ``scalar-tensor" mode. 

(b) Another difference is ``kernel function" $I_{\alpha\beta}$, coming from the different transfer functions between scalar and tensor perturbations. As shown in Eq.~(\ref{eq:II}), for ``scalar-scalar" mode, $I^2_{ss}\rightarrow \mathrm{const} $ in $u \rightarrow 0, v \rightarrow 1$ limit. However, for ``scalar-tensor" mode in Eq.~(\ref{eq:II}), $I^2_{st}\propto u^{-2}$ in $u \rightarrow 0, v \rightarrow 1$ limit, providing an enhancement factor related to small-momentum scalar perturbations.
    
In summary, the differences between $\Omega_\mathrm{gw}^{ss}$ and $\Omega_\mathrm{gw}^{st}$ at high frequencies result from distinct behaviors of the ``projection factors" $Q_{\alpha\beta}$ (related to geometry) and ``kernel function" $I_{\alpha\beta}$ (related to time evolution) in the limit $u\rightarrow0,\ v\rightarrow1$.

\end{widetext}

\end{document}